\documentclass[journal=jacsat,manuscript=article]{achemso}

\usepackage{chemformula} 
\usepackage[T1]{fontenc}

\author{Marcel Rey}
\affiliation{Physics Department, University of Gothenburg, 41296 Gothenburg, Sweden}
\email{marcel.rey@physics.gu.se}
\author{Giovanni Volpe}
\affiliation{Physics Department, University of Gothenburg, 41296 Gothenburg, Sweden}
\email{giovanni.volpe@physics.gu.se}
\author{Giorgio Volpe}
\affiliation{Department of Chemistry, University College London, 20 Gordon Street, WC1H 0AJ London, United Kingdom}
\email{g.volpe@ucl.ac.uk}

\title[An \textsf{achemso} demo]
  {Light, Matter, Action: Shining light on active matter}
\abbreviations{IR,NMR,UV}

\begin{document}

\begin{tocentry}

\includegraphics[width=8.25cm]{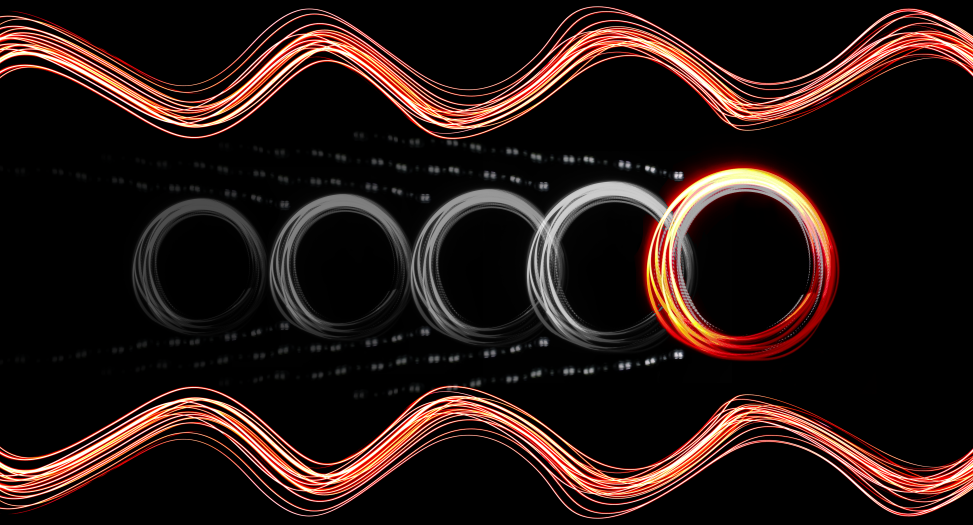}

\end{tocentry}

\begin{abstract}
    Light carries energy and momentum. It can therefore alter the motion of objects from atomic to astronomical scales.
    Being widely available, readily controllable and broadly biocompatible, light is also an ideal tool to propel microscopic particles, drive them out of thermodynamic equilibrium and make them active. Thus, light-driven particles have become a recent focus of research in the field of soft active matter.
    In this perspective, we discuss recent advances in the control of soft active matter with light, which has mainly been achieved using light intensity. We also highlight some first attempts to utilize light's additional degrees of freedom, such as its wavelength, polarization, and momentum. We then argue that fully exploiting light with all of its properties will play a critical role to increase the level of control over the actuation of active matter as well as the flow of light itself through it. This enabling step will advance the design of soft active matter systems, their functionalities and their transfer towards technological applications.
\end{abstract}

\section{Introduction}

In the last half century, the possibility of transporting and actuating objects with light has left the realm of science fiction to impact several fields of science and technology. Nowadays, a tremendous number of disciplines and applications benefit from actuating objects with light. These include optical manipulation \cite{volpe2022roadmap}, microfluidics \cite{Baigl2012photo}, nanomedicine \cite{Li2020}, manufacturing \cite{Han2016} and even space exploration \cite{Davoyan2021,volpe2022active}.

Light carries energy and momentum that can be transferred to materials via different types of light--matter interactions \cite{zemanek2019perspective}. While these effects are usually too small to be appreciated in our everyday life, their magnitude is big enough to influence the motion of microscopic objects\cite{jones2015optical}, whose energy fluctuations are comparable to the characteristic thermal energy $k_{\rm B}T$ with $k_{\rm B}$ the Boltzmann constant and $T$ the absolute temperature ($k_{\rm B} T \approx 4.14 \cdot 10^{-21}\,{\rm J}$ at room temperature). This is the realm of soft matter, i.e., the branch of science that studies systems and materials that can be deformed by relatively low energies on the order of thermal fluctuations.
Since the characteristic energy of a visible light photon is comparable to $k_{\rm B}T$, light is particularly well suited to interact with soft materials (e.g., a green photon has energy $E_{\rm photon} = h c / \lambda \approx 3.8 \cdot 10^{-19}\,{\rm J} \approx 90\, k_{\rm B}T$, where $h$ is the Planck constant, $c$ is the speed of light, and $\lambda=532\,{\rm nm}$ is the wavelength).

Active matter is a term used to include all \emph{living} and \emph{artificial} systems that can autonomously perform work for different tasks (e.g., move, transport cargoes, energy conversion) by utilizing energy available to them in their environment. These systems can develop rich forms of self-organization and collective dynamics\cite{vicsek2012collective}, leading to the emergence of complex properties, such as the possibility of interacting and evolving autonomously \cite{ramaswamy2017active,needleman2017active}.
At the macroscopic scale, examples of living active matter include animal groups and human crowds \cite{vicsek2012collective}, while active granular matter \cite{ramaswamy2017active} and robotic swarms \cite{dorigo2020reflections} represent their artificial counterparts. At the microscale (the focus of this perspective),  examples of living active matter include, e.g., bacterial cells \cite{wadhwa2021bacterial} and sperm cells \cite{gaffney2011mammalian}, while self-propelling colloids and microrobots are their man-made analogues \cite{bechinger2016active}. Systems at this scale are characterized by two main features: (1) Brownian fluctuations can influence these systems' motion and (2) inertia can be neglected \cite{purcell1977life}.

\begin{figure}[h!]
  \includegraphics[width=11cm]{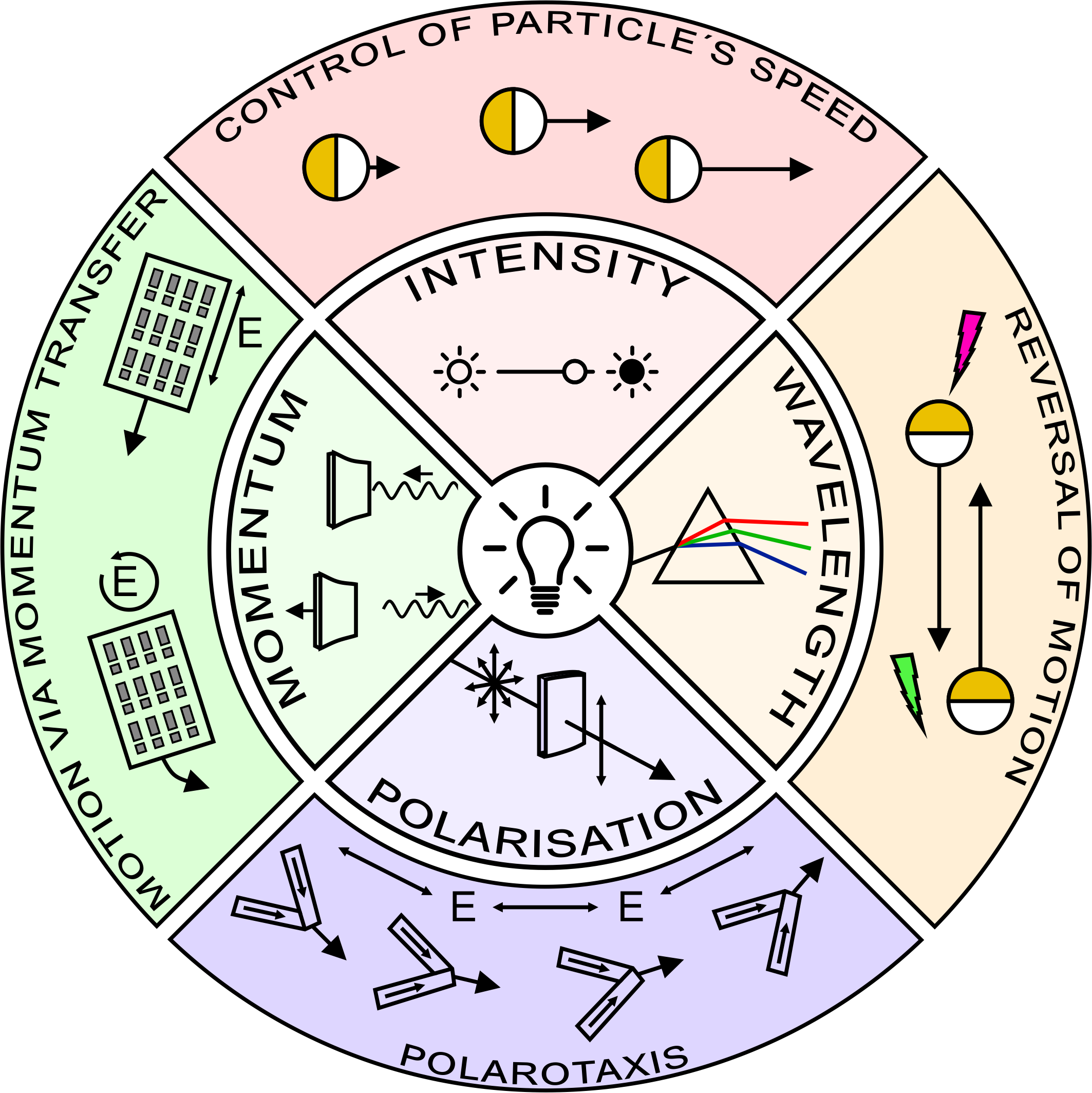}
  \caption{{\bf Actuation of active matter by different properties of light}. Different properties of light (intensity, wavelength, polarization and momentum) can be employed to control active matter. This schematics represents the main properties of light (inner circle) and some prominent examples of actuation (outer circle). (Top) The intensity of light typically correlates with the magnitude of the respective particle's propulsion mechanism \cite{buttinoni2012active}, so that the speed of the active particles can be adjusted by intensity. (Right) Different wavelengths can address different parts of heterogeneous active Janus particles, enabling control over propulsion direction and magnitude \cite{Vutukuri2020}. (Bottom) Nanomotors consisting of nanowires with a high dichroic ratio preferentially absorb polarized light, enabling polarotactic active movement controlled by the polarization state of the incident light.\cite{zhan2019strong} (Left) Light also carries momentum, which can propel matter: for example, microvehicles bearing metasurfaces that scatter light directionally can be accelerated via transfer of light momentum \cite{andren2021microscopic}.} 
  \label{Fig_OV}
\end{figure}

Being widely available, readily controllable and broadly biocompatible, light is an ideal tool to control microscopic particles and drive them out of thermodynamic equilibrium, thus making them active (Fig.~\ref{Fig_OV}). While the active matter community has enthusiastically adopted this tool to control microscopic active particles (e.g., bacteria, active colloids and active droplets), most studies have focused on exploiting the intensity of light, while neglecting the other degrees of freedom offered by light, such as its wavelength, its polarization and its linear and angular momenta (Fig.~\ref{Fig_OV}). Higher control will enable more fundamental scientific discoveries about far-from-equilibrium phenomena \cite{bechinger2016active}, while being also useful for applications, e.g., in sensing, nanomedicine and materials science \cite{sipova2020nanoscale}. Nowadays the prospects for light actuation and advanced particle tracking are ever brighter thanks to the development of several new technologies, such as cheaper lasers at all wavelengths, more versatile spatial light modulators, and higher-speed cameras.

In this perspective, we first discuss recent advances in the control of soft active matter actuation with light using its simplest degree of freedom (light intensity), with a focus on microscopic active systems such as microorganisms, active colloids and droplets. We then highlight first attempts and potential future mechanisms to utilize light's additional degrees of freedom, such as its wavelength, polarization and momentum. Finally, we propose potential avenues to increase the level of control over the actuation of soft active matter based on fully exploiting light's degrees of freedom.

\section{Active matter actuation by light intensity}

\begin{figure}[h!]
  \includegraphics[width=16.5cm]{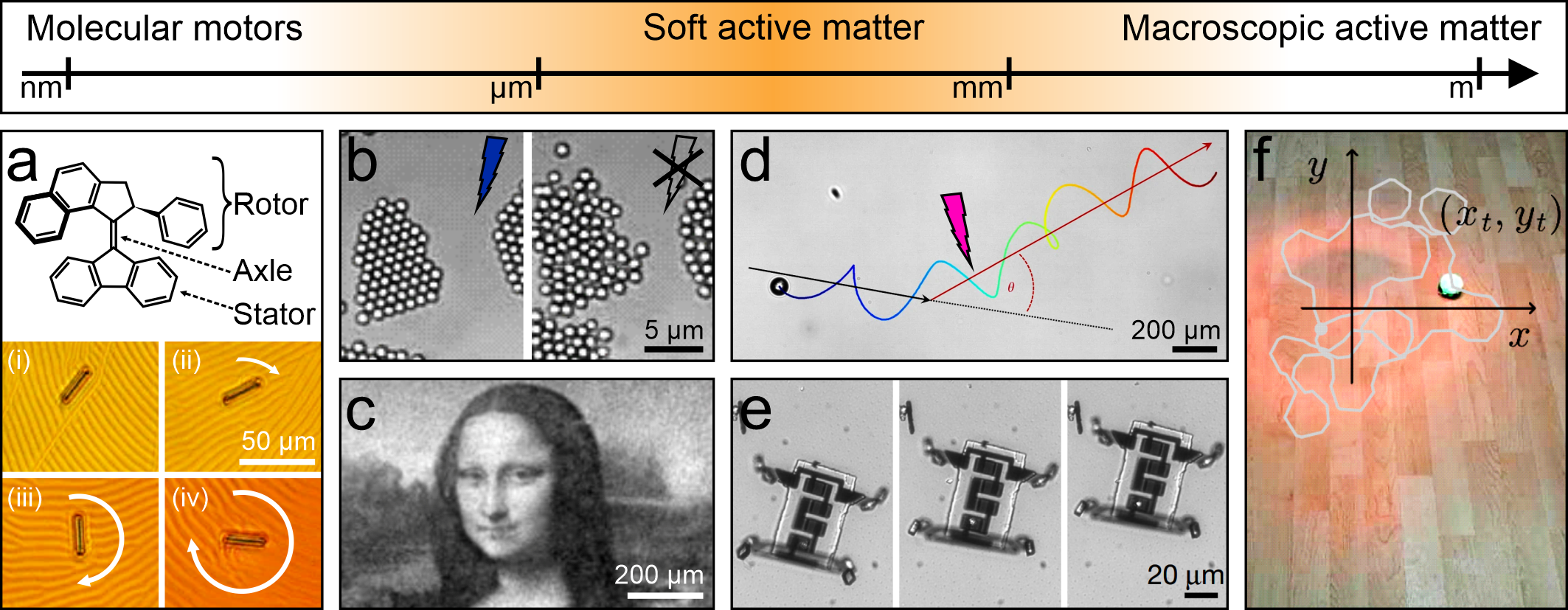}
  \caption{{\bf Active matter systems controlled by light intensity}. 
  Light intensity has been employed to control active matter systems at all length scales. a) On the molecular scale, light-responsive nanomotors can undergo a photochemical isomerization around the central double bond upon irradiation with UV light that results in helicity inversion (from right-handed to left-handed).\cite{Eelkema2006} 
  This motor is very effective at inducing helical organization in a liquid-crystal film, which can be harnessed to move microparticles placed on top of it ((i)-(iv)). Adapted with permission from ref \cite{Eelkema2006}. Copyright [2006] [Springer Nature]. b-e) Microscopic systems. b) light-activated Janus colloids self-organize into clusters under blue light but dissolve when the light source is turned off. Adapted with permission from ref \cite{palacci2013living}. Copyright [2013] [American Association for the Advancement of Science].
  c) Bacteria, genetically modified to swim smoothly with a light-controllable speed, can be arranged into complex and reconfigurable density patterns such as a portrait of Mona Lisa using a simple digital light projector. Adapted with permission from ref \cite{frangipane2018dynamic}. Copyright [2018] [eLife Sciences Publications Ltd].
  d) Self-propelling droplets of chiral nematic liquid crystals in surfactant-rich water propel in a screw-like motion. Photo-invertible chiral dopants allow converting between right-handed and left-handed trajectories upon UV irradiation. Adapted with permission from ref \cite{lancia2019reorientation}. Copyright [2019] [Springer Nature].
  e) Electronically-integrated micromotors consisting of a body containing standard silicon electronics and surface electrochemical actuator legs are able to walk by directing laser light to its photovoltaics that alternately bias the front and back legs. Adapted with permission from ref \cite{miskin2020electronically}. Copyright [2020] [Springer Nature]. 
  f) On the macroscale, phototactic robots can respond to light gradients, e.g., by adjusting their speed in response to the measured light intensity. Adapted with permission from ref \cite{mijalkov2016engineering}. Copyright [2016] [American Physical Society].
  } 
  \label{Fig_1}
\end{figure}

Thanks to the technological developments in the last decades, highly-controllable lasers and other light sources are nowadays easily available to most research labs. Light intensity is the most obvious light property that can be exploited to enable control over the actuation of microscopic active matter. This has been done at various scales, from molecular motors \cite{kassem2017artificial,credi2006artificial} (Fig.~\ref{Fig_1}a), to microscopic colloids, bacteria and droplets  \cite{bechinger2016active} (Fig.~\ref{Fig_1}b-d), to microrobots  \cite{palagi2018bioinspired,bunea2021light} (Fig.~\ref{Fig_1}e) and macroscopic robots\cite{mijalkov2016engineering}  (Fig.~\ref{Fig_1}f). In the following, we analyze how light intensity has been used to control the behavior of these systems, with an emphasis on systems at the micrometric scale (Fig.~\ref{Fig_1}b-e).

\subsection{Micro-organisms}

Several micro-organisms, including archea, bacteria and protists, have evolved to sense and respond to light. Phototaxis, whether towards (positive) or away from (negative) a light source, can be advantageous to optimize biological and physiological functions, such as photosynthesis, growth and the uptake of resources in competitive ecological contexts. For example, positive phototaxis can be beneficial for phototropic microorganisms, helping them to position and orient themselves to efficiently perform photosynthesis \cite{Menzel1979}. In these microorganisms, the response to light is usually mediated by photoreceptor proteins (e.g., proteorhodopsin and rhodopsin pigments) that are sensitive to light \cite{Menzel1979}. Most of these micro-organisms can measure light intensity gradients and move accordingly performing a biased random walk towards higher or lower light intensities. This is either achieved by probing changes in the signal over time \cite{McCain} or, in more complex micro-organisms, by directly measuring the gradient direction \cite{Kreimer2009,schuergers2016cyanobacteria}. A beautiful example of the latter is represented by the unicellular cyanobacterium \emph{Synechocystis} that can accurately and directly sense the position of a light source as the cell itself acts as a spherical microlens, allowing it to see the source and move towards it \cite{schuergers2016cyanobacteria}. The interaction among multiple phototactic micro-organisms can lead to the emergence of collective phenomena. For example, bioconvective flows form in systems of phototactic algae due to the formation of uneven mass distributions of the cells moving towards a light source \cite{dervaux2017light,arrieta2019light}. Time and space variations of the source lead to the dynamic triggering and reconfiguration of these bioconvective plumes \cite{dervaux2017light,arrieta2019light}. 

Beyond naturally photoresponsive microorganisms, the advent of optogenetics has enabled researchers to introduce exogenous DNA into non-photosensitive cells to express the production of light-sensitive proteins \cite{fenno2011development}. For example, scientists have engineered \emph{E. coli} bacterial cells to respond to red, green and blue light with the production of different pigments creating color photographs. \cite{fernandez2017engineering} Light-sensitive proteins have been also expressed in \emph{E. coli} to modulate their motility and consequently their population density by light \cite{walter2007light}, permitting the generation of dynamic bacterial patterns and images (Fig.~\ref{Fig_1}c) \cite{arlt2018painting,frangipane2018dynamic}. 

\subsection{Micromotors}

Inspired by these phototactic micro-organisms, various man-made self-propelling microscopic particles that can move in response to light have been developed (see, e.g., these recent reviews \cite{eskandarloo2017light,xu2017light,safdar2017light,chen2018light,wang2018light,villa2019fuel,sipova2020nanoscale,Palagi2019re}). 
In a homogeneous light-field, their motion can be described as a persistent random walk, similar to motile micro-organisms in homogeneous environments. 
In the presence of a light intensity gradient, their motion can become biased \cite{bechinger2016active}. 
A paradigmatic example of micromotors is constituted by Janus particles (named after the two-faced Roman god) \cite{howse2007self}. These are colloidal particles whose surface presents two different physico-chemical properties \cite{hu2012fabrication,Wang2022}. This asymmetry induces a local gradient in some thermodynamic properties (e.g., concentration, interfacial energy or temperature) across the particle that leads to its self-propulsion \cite{howse2007self,bechinger2016active}. Light can be used to create such an asymmetry across an illuminated particle (e.g., in its temperature profile or surface chemistry). One approach relies on coating one side of the particle with a photocatalytic material (such as platinum, palladium, hematite or titania) to locally decompose a chemical fuel (usually hydrogen peroxide) in water and create a local concentration to drive the particle’s self-diffusiophoresis \cite{ibele2009schooling, hong2010light,solovev2011light,palacci2013living} or self-electrophoresis \cite{dai2016programmable,Dong2016,Du2020}.  Alternatively, light absorption in Janus particles half-coated with a light-absorbing material (e.g., gold or carbon) can also lead to self-propulsion directly or indirectly due to the formation of a local temperature gradient across the particle because of selective heating at the absorbing side \cite{jiang2010active,Volpe2011_pattern,buttinoni2012active,qian2013harnessing,shao2020photoactivated,dietrich2020microscale,Wang2022}. Differently from their biological counterparts which can move by body deformation, most of these synthetic micromotors have rigid shapes and only recently light-responsive reconfigurable microswimmers have been proposed \cite{alvarez2021reconfigurable,Zhang2019,Magdanz2014,Mourran2017,vankesteren}. 
As in the case of micro-organisms, artificial Janus particles can also orient in the light-field and feature a biased directional phototactic behavior in a light intensity gradient \cite{buttinoni2012active,lozano2016phototaxis}. 

While at the individual particle level these artificial micromotors have found interest as a promising route to develop novel applications in nanomedicine and environmental remediation \cite{safdar2017light,wang2018light}, complex collective behaviors have also been reported when these individual units self-organize into larger light-activated clusters, including the formation of living crystals (Fig.~\ref{Fig_1}b) \cite{palacci2013living}, inverse crystallization\cite{Huang2020}, crystal annealing,\cite{Ramananarivo2019} different lattice structures \cite{singh2017non}, active colloidal molecules \cite{Schmidt2019,grauer2021active,Madden2022}, dynamic pattern formation \cite{ilday2017rich,makey2020universality,Massana-Cid2018}, metamachines \cite{aubret2018targeted} and even functional photonic materials \cite{trivedi2022self}.
These emerging behaviors are interesting as model systems to study self-organization in living matter but also as a novel route to develop next-generation materials. Mostly, these complex collective behaviors are governed by physical forces such as steric, phoretic and hydrodynamic interactions. Recently, however, light has been used to encode more complex behaviors which also include feedback interactions among the units \cite{khadka2018active,lavergne2019group,muinos2021reinforcement}.

\subsection{Active Droplets}

Droplets are small volumes of liquid separated from their surroundings by at least one interface \cite{malinowski2020advances}. Because of their small size, they can be used as versatile transport vessels and reactors in microfluidics for applications in chemistry, biology, and nanomedicine \cite{malinowski2020advances}. Active droplets are a particular class of droplets that can either self-propel in isolation or do so in response to other neighboring droplets in an emulsion \cite{maass2016swimming,birrer2022we}. The main physical effect that induces these droplets' motion is the Marangoni effect, where a gradient of surface tension drives mass transport towards areas of higher surface tension. In the case of active droplets, the Marangoni effect is (self-)induced by the droplet itself or by surrounding droplets. Such gradients of surface energy can be produced optically, e.g., by illuminating the droplet surface and harnessing the thermal or photochemical effects of the light absorbed within the droplet \cite{Baigl2012photo,ryazantsev2017thermo,birrer2022we,Kawashima2017}.  For example, lipophilic droplets stabilized by photoresponsive surfactants can move in light gradients. Light irradiation induces the dissociation of photoresponsive surfactants combined with a rapid pH change in the surrounding aqueous phase, which results in fast movement of the droplet away from the light source due to a change in surface tension \cite{florea2014photo}.
In self-propelling liquid crystal droplets containing photo-invertible chiral dopants, light irradiation allows converting between right-handed or left-handed screw-like trajectories (Fig.~\ref{Fig_1}d).\cite{lancia2019reorientation}
Recently, similar light-induced motility effects away or towards a light source were also reported in droplets of various materials with the addition of different photoresponsive molecules, such as surfactants\cite{suzuki2016phototaxis,kaneko2017phototactic}  and inorganic particles \cite{singh2020interface}.

\subsection{Microrobots}

The possibility of fabricating active particles with complex functionalities has driven the development of novel microrobots, i.e., robots with characteristc sizes below 1 mm \cite{palagi2018bioinspired,zeng2018light_rev}. While these systems are at the very edge of what is typically considered soft active matter, they are a powerful reminder of what can be achieved by light actuation. Due to their relatively larger size, inertial effects, rather than viscous, are more prominent in determining their motion, and Brownian fluctuations less relevant \cite{lowen2020inertial}. In particular, liquid crystalline elastomers (LCE) are a common material employed to realize biomimetic micromotors capable of autonomous locomotion in response to light \cite{jiang2013actuators}. Beyond the realization of devices at the millimeter scale \cite{rogoz2016light,gelebart2017making,zeng2018light,shahsavan2020bioinspired,cheng2022light}, these materials have been successfully employed to realize microrobots in the submillimeter range, such as walkers on solid surfaces \cite{zeng2015light} and biomimetic swimmers in fluids \cite{palagi2016structured}. Similar shape-changing walkers and grippers were realized based on photo-sensitive spiropyran-based hydrogels.\cite{Francis2017,li_gripper} Recently, a new class of microrobots has been demonstrated that integrated electronic components with light actuation, thus providing a stepping stone towards mass-manufacturing silicon-based functional robots at the microscopic scale (Fig.~\ref{Fig_1}e) \cite{miskin2020electronically}. 

\section{Actuation by other properties of light}

Differently from intensity, other light properties have not been extensively exploited for the actuation of soft active matter yet. In this section, we will briefly review how they have been used so far in this context (with a focus on wavelength, polarization and transfer of momentum) and what further possibilities they offer.

\subsection{Wavelength}
The wavelength determines the color of the light as well as the energy carried by each photon. Optimal light conditions crucial for microorganism, e.g. cyanobacteria,\cite{Nakane} that grow by capturing energy from sunlight. They demonstrate positive phototaxis towards green light as it is their preferred energy source for oxygenic photosynthesis, while they show negative phototaxis away from strong light or UV light as it causes cell damage.\cite{Nakane}
Similarly, marine zooplankton \textit{Platynereis} larvae exposed to UV light swim downwards, away from the light source, while cyan light makes the larvae swim in the reverse, upwards direction.\cite{Veraszto} In the ocean, UV light is most intense near the surface, while cyan light reaches greater depths. \textit{Platynereis} larvae may thus use the ratio between UV and cyan light as a ``depth gauge'' during vertical migration.\cite{Veraszto}

In artificial active matter, wavelength is the second most important and exploited property of light after intensity. The wavelength is often a boundary condition imposed by the materials employed in the experiment. For example, Janus particles with metallic caps can be heated by light of a specific wavelength depending on the cap's material, e.g., green matches the plasmon resonance of gold ($\lambda\approx530\,{\rm nm}$), blue that of silver ($\lambda\approx400\,{\rm nm}$), and UV that of platinum ($\lambda\approx260\,{\rm nm}$).

\begin{figure}[t!]\includegraphics[width=16.5cm]{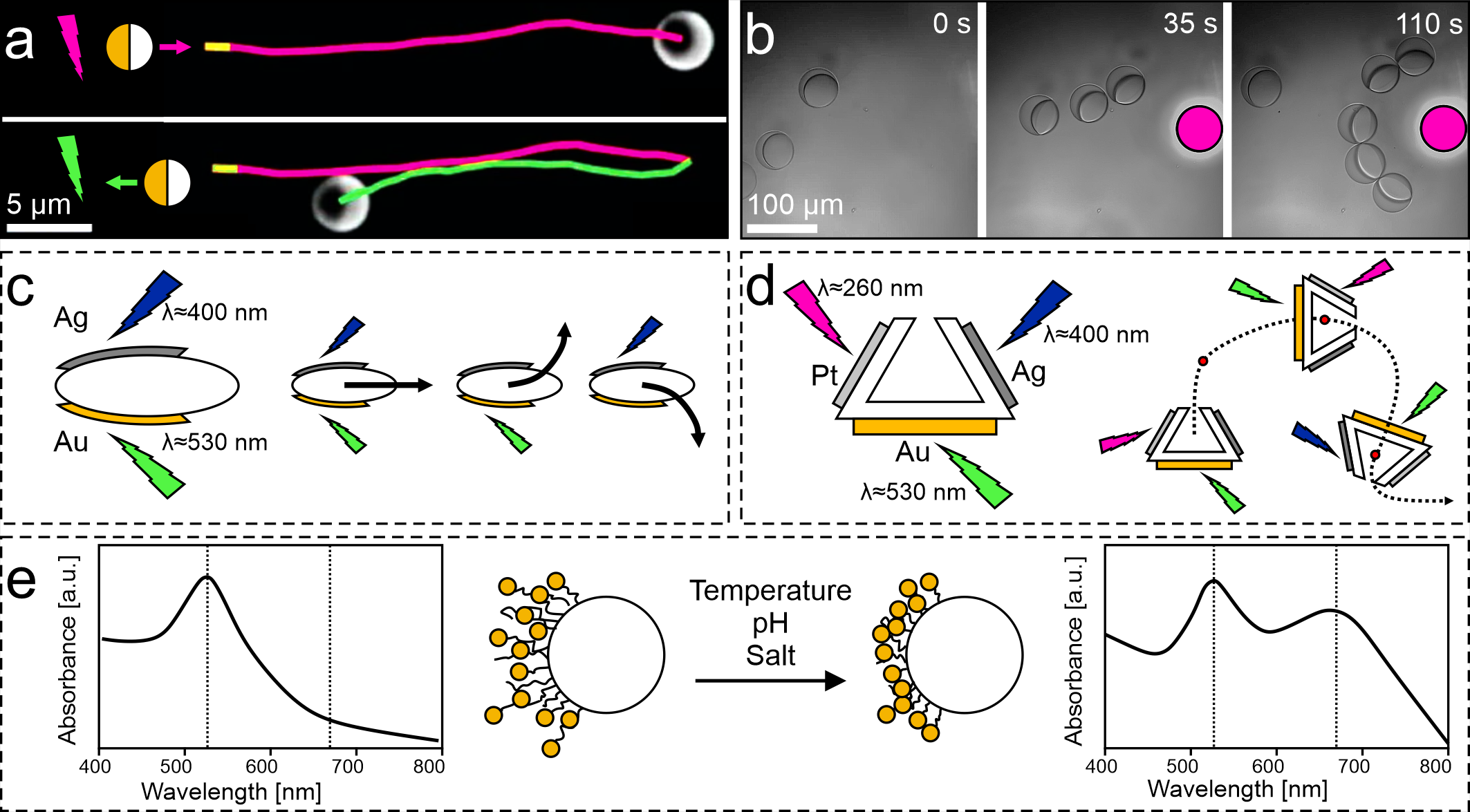}
  \caption{{\bf Active matter systems controlled by wavelength}.
  a) Forward trajectory of Au-coated anatase TiO$_2$ Janus particles in a H$_2$O$_2$ solution upon illumination with UV light (top, magenta trajectory) and reverse direction upon illumination with green light (bottom, green trajectory).  Adapted with permission from ref \cite{Vutukuri2020}. Copyright [2020] [Springer Nature].
  b) AzoTAB-stabilized Janus emulsions under bright-field blue light irradiation  self-assemble towards a localized UV light spot (filled circle). Adapted with permission from ref \cite{Frank2022}. Copyright [2020] [Springer Nature].
  c-e) Potential future uses of wavelength as control strategy for active particles. 
  c) Elongated particles (e.g., ellipsoids) with two different metal patches can be steered using two wavelengths of light. 
  d) U-shaped particle with three metal patches can work as cargo carriers, with full control over their planar movement.  
  e) In stimuli-responsive polymer brush coated Janus particles decorated with plasmonic nanoparticles, external stimuli such as temperature, pH or salt concentration can collapse the polymer and shift the absorbance spectrum,\cite{Christau2017} affecting the particle mobility.
   } 
  \label{Fig_2}
\end{figure}

Multiple wavelengths have been combined in a single experiment to control the propulsion of different types of active particles\cite{Zheng2017_2} (or different parts of an active particle) to achieve more complex particle's behaviors.\cite{Vutukuri2020,Wang2018,Jang2017,Sridhar2020} 
For example, TiO$_2$ Janus particles with either cobalt oxide caps\cite{Sridhar2020} or metal caps\cite{Vutukuri2020,Wang2018,Jang2017} combine a complex interplay between adsorption of light at different wavelengths and the respective catalytic and photochemical processes occurring on each side of the particle. Adjusting the wavelength of light enables control over the propulsion direction (including its on-demand reversal) \cite{Vutukuri2020,Wang2018} and magnitude\cite{Vutukuri2020,Jang2017,Wang2018,Sridhar2020} (Fig.~\ref{Fig_2}a). Similarly, hybrid active particles made from two different photocatalysts, e.g. TiO$_2$ and CuO$_2$, catalyze hydrogen peroxide over differing ranges of wavelength,\cite{ONeel-Judy2018} which can lead to a wavelength-dependent translational and rotational swimming behavior.\cite{ONeel-Judy2018} 
Alternatively, photoelectrochemically driven nanotree microswimmer loaded with photosensitizer dyes can be driven and steered by visible light using different wavelengths.\cite{Zheng2017_2} 

Strategies to manipulate liquid interfaces and droplets typically employ azobenzene-derived surfactants such as AzoTAB.\cite{azotab}
These molecules can be reversibly switched between two conformations of different polarity by subsequent illumination with UV (365 nm) and blue (475 nm) light, which affects the interfacial tension of interfaces stabilized by the surfactant. For example,
this principle enabled the omnidirectional manipulation of oil droplets\cite{diguet2009photomanipulation} and liquid marbles\cite{Kavokine2016} floating on the surface of an aqueous solution of AzoTAB surfactants. By changing the wavelength of the light used to illuminate the edge of the droplets or liquid marbles, it was possible to reversibly repel them from the incident beam (UV illumination) or attract them towards it (blue illumination).\cite{diguet2009photomanipulation,Kavokine2016} 
Similarly, oil droplets can be propelled in the proximity of azobenzene-stabilized micelles.\cite{Ryabchun2022}
Changing the wavelength of the light induces a change in micelle geometry, which impacts the movement pattern of
the droplets, resulting in a run-and-halt behavior.\cite{Ryabchun2022}
Further, Janus emulsions stabilized with AzoTAB under blue light irradiation move towards a UV light spot around which they self-assemble in an ordered fashion (Fig.~\ref{Fig_2}b).\cite{Frank2022} Finally, fatty acid droplets containing photo-sensitive spiropyran can move towards visible light sources and away from UV light sources, thus enabling their manipulation in three dimensions.\cite{xiao2018moving}

Potential future uses of the wavelength as a control mechanism of an active system may include:

\begin{itemize}
    \item {\bf Multiple resonant shapes}. Complex active particle shapes can be designed to respond to different wavelengths, e.g., by exploiting characteristic plasmonic resonances (hence wavelength-selective enhanced light absorption) of different metals. For example, ellipsoidal or rod-shaped anisotropic Janus particles with two different metal patches instead of one\cite{Kirvin2021} would enable addressing each individual patch or both simultaneously by adjusting the wavelength of light. The anisotropic nature of the particle should allow control over the direction and magnitude of propulsion, where particles can either move straight  or rotate clockwise or counterclockwis depending on the light wavelength and intensity (Fig.~\ref{Fig_2}c). The same underlying principle can be used to design even more complex units, e.g., U-shaped particles with multiple plasmonic patches, which could additionally allow the reversal of the direction of motion for loading/unloading cargoes (Fig.~\ref{Fig_2}d). 
    
    \item {\bf Stimuli-responsive plasmonic or photonic resonance shifts via elastic deformations}. 
     The elastic deformation of stimuli-responsive polymer brushes decorated with plasmonic nanoparticles has been employed to tune the wavelength of the light absorbed by this composite structure: external stimuli such as changes in pH,\cite{Tokareva2004} temperature\cite{Christau2016} or salt concentration \cite{Christau2017} can collapse the polymer brush and bring the plasmonic nanoparticles in closer contact, thus red-shifting the absorption spectrum. We suggest to employ the same concept for active particles, which will then be able to adapt their activity to their local environment (Fig.~\ref{Fig_2}e). 
    A further, though slightly more futuristic, approach is inspired by the skin of chameleons and cephalopods: these animals can camouflage by actively tuning the photonic response of their skin through elastic deformation, thus changing the wavelength of the reflected and absorbed light. Similarly, photonic active particles could be realized with stimuli-responsive hydrogel opal films,\cite{Galei} where the photonic band gap can be adjusted via external stimuli such as temperature.
\end{itemize}
\subsection{Polarization}

Polarization is the property of light waves which describes the oscillation of the electric field in the direction perpendicular to the wave propagation. For example, light from the sun is unpolarized, i.e., there is no preferred orientation for this oscillation. Unpolarized light can become polarized when it is scattered or passes through polarizing filters that select only certain orientations of the electric field.  
In linearly polarized light, the electric field oscillates in a single direction perpendicular to the propagation direction. In circularly polarized light, the field rotates at a constant rate around the direction of propagation as the wave travels. 

\begin{figure}[h!]
  \includegraphics[width=16.5cm]{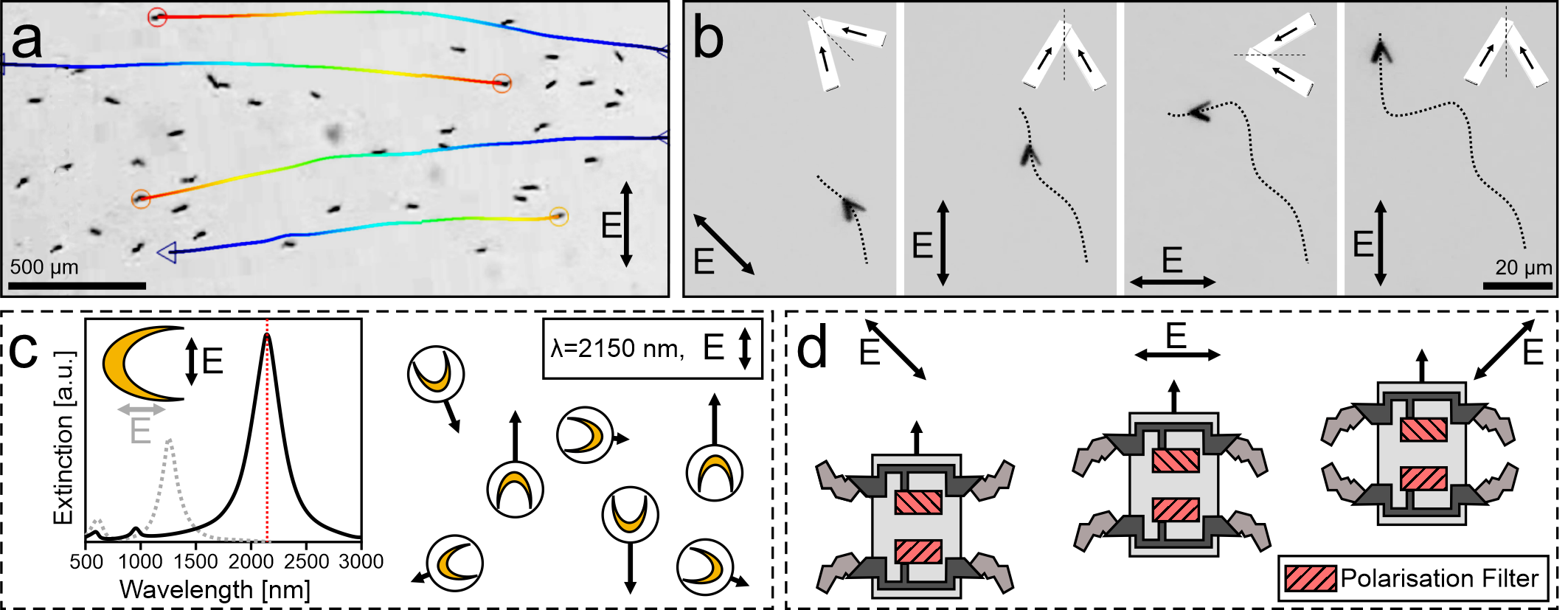}
  \caption{{\bf Active matter systems controlled by polarization}. 
  a) Polarotaxis  in photoresponsive algae (\emph{Euglena gracilis}) leads to movement perpendicular to the polarization of light. Adapted with permission from ref \cite{Yang2021}. Copyright [2021] [American Physical Society]. 
  b) Nanomotors consisting of nanowires with a
high dichroic ratio preferentially absorb polarized light, enabling polarotactic active movement and steering controlled by the polarization state of the incident light. Adapted with permission from ref \cite{zhan2019strong}. Copyright [2019] [John Wiley and Sons].
  c,d) Potential future uses of polarization as control mechanism in active particles. c) Plasmonic nanocrescents  feature polarization-dependent resonances and near-field enhancement at their tips.\cite{Goerlitzer2020} For a fixed wavelength, the propulsion strength of active particles driven by similar nanostructures therefore would depend on their orientation relative to the polarization of light, leading to predominant motion in the direction where the absorption of a given polarization is stronger. 
  d) Electronically-integrated micromotors\cite{miskin2020electronically} equipped with polarizing filters in front of photovoltaic components could allow the polarization-dependent control of specific actuators to steer the particle’s self-propulsion with the light polarization.
  } 
  \label{fig_3}
\end{figure}

Sensitivity to polarization is not uncommon in nature. For example, many insect species bear photoreceptors in a small dorsal rim area of the eye that detect polarized skylight to improve their navigation skills.\cite{Mathejczyk2017} Furthermore, polarization sensitivity helps squids detect transparent, yet polarization-active zooplankton under partly polarized light.\cite{Shashar} The vision of the mantis shrimp holds the world record for the most complex visual system: these marine crustaceans have up to 16 photoreceptors and can see UV, visible and polarized light; they are also the only animal known to detect circularly polarized light, which may serve as a secret communication system.\cite{Gagnon2015}
At the microscale, \emph{Euglena gracilis} cells (an alga) exhibit polarotaxis behavior, which aligns their motion direction perpendicular to the light polarization.\cite{Hader1987} Their polarotaxis can also be used to guide the collective movement of these algae (Fig.~\ref{fig_3}a).\cite{Yang2021}

In the manmade world, polarized light has been widely exploited in optical manipulation to control the orientation and rotation of particles via transfer of linear and angular momentum.\cite{zemanek2019perspective} However, only recently, nanomotors with a high dichroic ratio have been shown to respond to polarized light. These nanomotors are constituted of nanowires with a ${\rm ZnO}$ shell and a ${\rm Sb_2Se_3}$ core, whose anisotropic crystal structure preferentially absorbs light polarized along the wire, hence enhancing their self-propulsion speed. \cite{zhan2019strong} By connecting two cross-aligned dichroic nanowires, the authors of this work were able to realize artificial polarotactic active particles, whose navigation can be controlled by the polarization state of the incident light. \cite{zhan2019strong} 

However, when it comes to actuating artificial active matter, polarization is still under-explored. 
Examples of possible future uses of polarization to control the motion of active particles are:

\begin{itemize}
    \item {\bf Polarization-dependent absorbers}. The scope for active particles whose self-phoretic forces depend on the absorption of specific polarizations of light is broad. This can be achieved employing dielectric structures (as in ref.~\cite{zhan2019strong}) or metallic structures, e.g., the plasmonic nanocrescents in Fig.~\ref{fig_3}c, which feature multiple polarization-dependent resonances combined with near-field enhancement at their tips.\cite{Goerlitzer2021,Goerlitzer2020array} The propulsion speed of active particles featuring similar nanostructures would therefore depend on their orientation relative to the polarization of light. For example, in Fig.~\ref{fig_3}c, this would lead to an enhanced propulsion of the particles in the direction parallel to the polarization of light, thus leadign to a polarotactic behavior as the algae cells in Fig.~\ref{fig_3}a.\cite{Yang2021}
    \item {\bf Polarized photovoltaics}. The integration of electronic components in microrobots\cite{miskin2020electronically} and metavehicles\cite{andren2021microscopic} could be exploited to generate polarization-dependent motion. The electronic components could be simple circuits made from standard inorganic or organic photovoltaics and metal interconnects powering some actuators on the active particle. \cite{miskin2020electronically} The use of polarizing filters in front of the photovoltaic components could allow the polarization-dependent control of specific actuators to steer the particle's self-propulsion with the light polarization (Fig.~\ref{fig_3}d). 
\end{itemize}

\subsection{Transfer of momentum}

The existence of light momentum is foundational to the whole field of optical trapping and optical micromanipulation \cite{jones2015optical,Marago2013,zemanek2019perspective}.
When light interacts with matter, the change in linear and angular optical momentum induces forces and torques. 
In most cases, these forces and torques are used to hold particles in place or to generate some deterministic, controllable motion. There are nevertheless some cases where they have been employed to alter the random motion of microscopic particles in interesting ways. For example, the forces produced by random light fields have been employed to alter the diffusion of Brownian particles \cite{Volpe2014_opt,Volpe2014,Bewerunge2016,Pastore2021,Pastore2022,Segovia-Gutierrez2020,Zunke2022} leading also to superdiffusive behavior in the presence of time-varying patterns.\cite{Douglass2012,Volpe2014,Bianchi2016}

\begin{figure} [h!]
  \includegraphics[width=16.5cm]{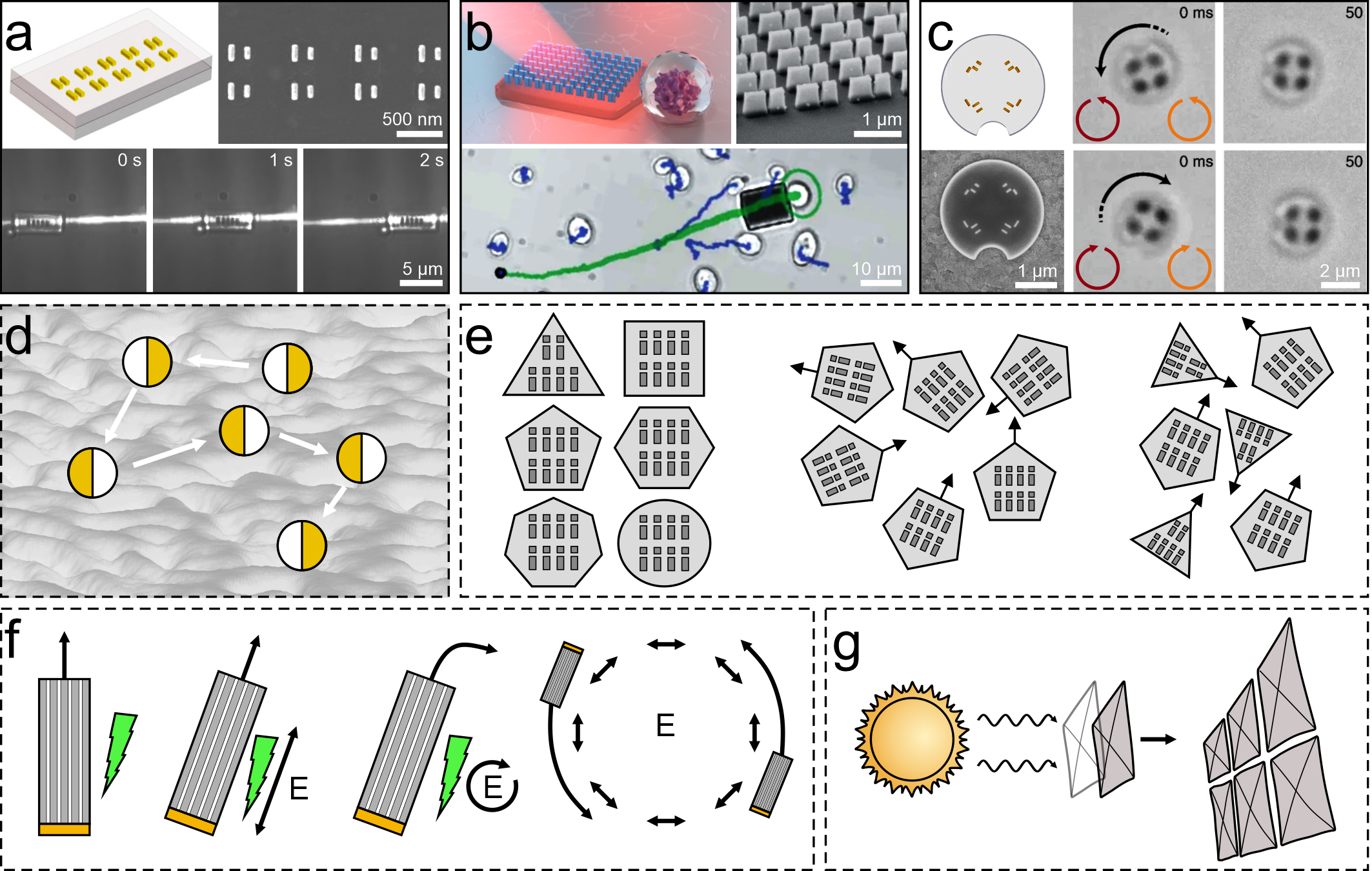}
  \caption{{\bf Active matter systems controlled by transfer of light momentum}. 
  a) Plasmonic linear nanomotor driven by momentum transfer of light. Adapted with permission from ref \cite{tanaka2020plasmonic}. Copyright [2020] [American Association for the Advancement of Science].
  b) Microvehicles containing a directional scattering metasurface. The transfer of momentum leads to a straight propulsion under linearly polarized light and circular motion under circularly polarized light. These metavehicles can be used also to move some microorganisms present in the solution (bottom panel). Adapted with permission from ref \cite{andren2021microscopic}. Copyright [2021] [Springer Nature].
  c) Light-driven microparticles containing four chiral plasmonic resonators are maneuvered by adjusting the optical power for each resonator using two overlapping unfocused light fields at 830 nm (orange arrow) and 980 nm (red arrow) with right- and left-handed circular polarization, respectively. Adapted with permission from ref \cite{Wu2022}. Copyright [2022] [Springer Nature].
  d-g) Potential future uses of momentum transfer in active matter. 
  d) Active particles either propelled by light or by chemical fuels can explore a speckle light pattern according to some non-trivial random motion statistics (e.g., by a Fickian yet non-Gaussian diffusion\cite{Pastore2021}).
  e) Active momentum-driven particles with different shapes can generate emergent collective self-assembly behaviors, as already theoretically modeled~\cite{Moran2022,Moran2022_2}. 
  f) Active birefringent Janus particles with a metal cap on one side\cite{tang_birefr} can combine the propulsion of Janus particles with the orientation in polarized light of birefringent particles. Such particles would move along the polarization direction of linearly polarized light or show circular motion in circularly polarized light.
  g) Solar sails\cite{Davoyan2021} propelled by the light of the sun could self-assemble in space into complex devices, e.g., space telescopes.
  } 
  \label{Fig_6}
\end{figure}

Within the field of active matter, recent work exploited the transfer of momentum using unfocused light to propel microscopic vehicles with incorporated plasmonic or dielectric metasurfaces that generate lateral optical forces due to directional light scattering along the side of the structure (Fig.~\ref{Fig_6}a-c).\cite{tanaka2020plasmonic,andren2021microscopic,Wu2022} For example, microvehicles with directional light-scattering nanostructures arranged in parallel can propel forward upon illumination with linearly polarized light due to transfer of linear momentum (Fig.~\ref{Fig_6}a,b)\cite{tanaka2020plasmonic,andren2021microscopic}, and they can be steered left or right by circularly polarized light thanks to transfer of angular momentum (Fig.~\ref{Fig_6}b).\cite{andren2021microscopic} As an example of application, these metavehicles were also employed for the micromanipulation of colloidal particles and micro-organisms (Fig.~\ref{Fig_6}b).\cite{andren2021microscopic} Alternately, rotation under plain linearly polarized light can be achieved by arranging the scatterers in a circle (Fig.~\ref{Fig_6}c).\cite{tanaka2020plasmonic,Wu2022} Moreover, microvehicles with four individually addressable chiral plasmonic nanoantennas acting as nanomotor enable full motion control in two dimensions in all three independent degrees of freedom (two translational and one rotational):\cite{Wu2022}  Similar to macroscopic drones but in two dimensions, these microvehicles are maneuvered by adjusting the optical power for each nanomotor using two overlapping unfocused light fields at $\lambda = 830\,{\rm nm}$ and $\lambda = 980\,{\rm nm}$, each with right- or left-handed polarization (Fig.~\ref{Fig_6}c).\cite{Wu2022}

Further possible uses of optical forces and torques for the field of active matter are the following:

\begin{itemize}
    \item {\bf Complex optical fields.} The use of complex light fields (e.g., random speckle light fields) is a promising way to generate non-trivial optical potentials that can influence the individual and collective motion behavior of active particles (Fig.~\ref{Fig_6}d). For example, the motion of non-light-driven (e.g., catalytic) active particles within random light fields may show a competition between their propulsion activity and the retardation introduced by the speckle field as function of increasing light intensity, which has been predicted in theoretical work.\cite{Paoluzzi2014} On the other hand, active particles that are driven by light may be accelerated in speckle fields leading to the emergence of superdiffusive patterns. 
    \item {\bf Complex metavehicles.} Metavehicles propelled by momentum transfer can be fabricated in any shape without interfering with their propulsion mechanism (Fig.~\ref{Fig_6}e). This makes them a promising model system to study collective phase behaviors of active particles as a function of particle shape, which has recently been theoretically modeled. \cite{Moran2022,Moran2022_2} 
    \item {\bf Birefringent active particles.} Birefringent particles, e.g., metamaterial nanoparticles,\cite{tang_birefr} containing an absorbing cap on one side would combine the propulsion of Janus particles with the re-orientation capabilities in polarized light of birefringent particles\cite{tang_birefr} (Fig.~\ref{Fig_6}f). Such hybrid particles can be expected to propel parallel to the polarization of light but also feature circular motion under circularly polarized light due to transfer of angular momentum.
    \item {\bf Active matter in space.} Self-organization of multiple active particles under the action of optical forces can find fruitful applications in space exploration. For example, several solar sails\cite{Davoyan2021} might interact through optical binding to generate complex collective behaviors and to produce large self-organized devices: e.g., new space telescopes with effective lenses made by self-organized active particles which might be much larger than the Webb telescope (Fig.~\ref{Fig_6}g). 
\end{itemize}

\section{Further actuation by structured light}

\begin{figure} [p]
  \includegraphics[width=16.5cm]{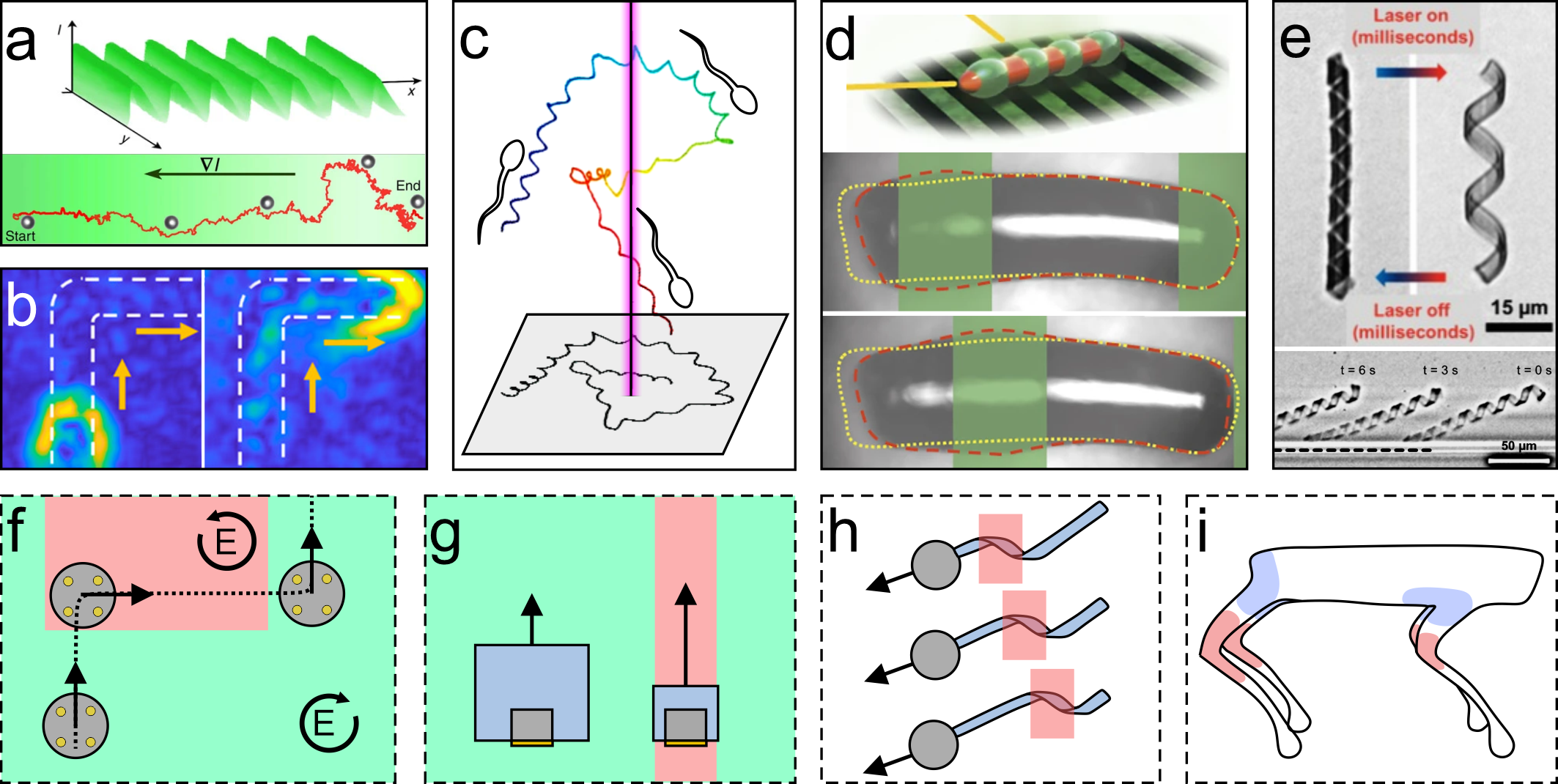}
  \caption{{\bf Active matter systems interacting with structured light}. 
  a) Janus particles in a critical mixture of water–lutidine align such that they move along the gradient of light toward low light intensities. Directed particle transport over arbitrarily long distances can then be achieved using periodic saw-tooth-like light profiles. Adapted with permission from ref \cite{lozano2016phototaxis}. Copyright [2016] [Springer Nature].
  b) Structured light can guide the motion waves among photochemically-activated colloids. Adapted with permission from ref \cite{Chen2022}. Copyright [2022] [American Physical Society].
  c) Structured UV light can create spatio-temporal chemical landscapes by releasing caged chemo-attractants, which guide the movement of sperms. Adapted with permission from ref \cite{Jikeli2015}. Copyright [2015] [Springer Nature].
  d) Spatiotemporally structured light induces intra-body shape changes in microrobots consisting of photoactive liquid-crystal elastomers, which enables self-propulsion by generating a traveling-wave motion. Adapted with permission from ref \cite{palagi2016structured}. Copyright [2016] [Springer Nature].
  e) Temporally structured light enables rapid dynamic switching between the configurations of helical composite hydrogel microrobots, enabling translational movement near a solid surface. Adapted with permission from ref \cite{Mourran2017}. Copyright [2016] [John Wiley and Sons].
  f-i) Potential future uses of structured light to actuate and control active matter.
  f) Metavehicles are able to change their motion direction (liner or circular) depending on the polarized polarization of the illuminating light.\cite{andren2021microscopic,Wu2022} Structured light landscapes with different local polarization could guide the movement of such microvehicles.
  g) Microrobots could comprise temperature-responsive bodies that shrink upon irradiation with IR light, thus reducing their drag force and increasing their propulsion magnitude. Structured light with different local wavelengths could spatiotemporally change their propulsion magnitude.
  h) Spatiotemporally structured light could locally bend hydrogel nanoribbons\cite{Mourran2017,Zhang2017scr,palagi2016structured} and induce a snake-like motion, which could be exploited to propel microparticles. 
  i) Microwalkers could be driven by hydrogel-gold nanoparticle composites, which serve as artificial muscles and joints in response to light. 
  Using spatiotemporally structured light, each artificial element could be addressed individually to enable microscale artificial walking.
   } 
  \label{Fig_7}
\end{figure}

A more holistic approach to the use of light for active matter will entail the full control in space and time of its properties, including amplitude, phase, polarization and momentum.\cite{Angelsky2020,Rubinsztein-Dunlop2017} 
Simple forms of structured light have already been employed in some active matter experiments.
For example, the fact that Janus particles in a critical mixture of water–lutidine feature negative phototaxis in light gradients by drifting towards lower light intensities because of diffusiophoretic torques (Fig.~\ref{Fig_7}a) has been exploited in saw-tooth-shaped static
light profiles to make particles undergo directed motion over arbitrarily long distances (Fig.~\ref{Fig_7}a).\cite{lozano2016phototaxis}
Furthermore, structured light has been used to guide traveling motion waves among photochemically-activated oscillating colloids.\cite{Chen2022} Silver chloride (AgCl) particles in dilute hydrogen peroxide solutions under UV light illumination  exhibit both single-particle and collective oscillations in their motion, which arise due to an oscillatory, reversible conversion of AgCl to silver metal at the particle's surface.\cite{Ibele2010ag} These motion waves can be guided by spatial light patterns, thus enabling a precise and programmable control over the motion waves' origin, path and direction (Fig.~\ref{Fig_7}b).\cite{Chen2022} 
In biological systems,  
structured UV light enabled the creation of defined 3D spatio-temporal chemical landscapes by releasing caged chemoattractants, which were used to investigate the chemotactic navigation mechanism of sperm (Fig.~\ref{Fig_7}c).\cite{Jikeli2015} Further, \emph{Euglena gracilis} algae swim in polygonal trajectories when exposed to a sudden increase in light intensity.\cite{Tsang2018} In spatially structured light landscapes with different light intensities, algae coming from low light to high light intensity start polygonal swimming or localized spinning, making the cells turn around.\cite{Tsang2018}

Structured light can also be used to induce body deformation in microswimmers leading to their motion. For example, spatiotemporally structured light based on interference patterns was used to power and control intra-body shape changes in microrobots consisting of photoactive liquid-crystal elastomers,\cite{palagi2016structured} which were able to self-propel by generating a traveling-wave motion (Fig.~\ref{Fig_7}d).\cite{palagi2016structured,VonRohr2018} 
Temporally structured light was further used to actuate soft microrobots. These microrobots consist of a temperature-responsive hydrogel filled with gold nanorods that enable fast heating/cooling dynamics upon irradiation with IR light combined with volumetric shape changes.\cite{Mourran2017} Covering the hydrogel composite on one side with a thin gold layer restricts the swelling and shrinking, leading to the formation of helical configurations.\cite{Mourran2017,Zhang2017scr,Mourran2021} Temporally structured light enables rapid dynamic switching between left-handed and right-handed helical configurations\cite{Zhang2017scr} as well as translational movement near a solid surface (Fig.~\ref{Fig_7}e).\cite{Mourran2017}

As the possibilities for structuring light increase with the advancement of light modulating devices, control of active matter with structured light could include:

\begin{itemize}

    \item {\bf Control of microvehicles by structured polarization.} Structured light with different local polarizations or wavelengths could guide the movement of microvehicles\cite{andren2021microscopic,Wu2022} that can change their motion patters depending on the polarization of the illuminating light (Fig.~\ref{Fig_7}f). 

    \item {\bf Shrinkable microrobots in structured intensity fields.} Structured light could spatiotemporally change the propulsion magnitude  of 
    microrobots comprising temperature-responsive bodies that shrink upon illumination with infrared light,\cite{Mourran2017,vankesteren,alvarez2021reconfigurable} thus reducing their drag force and increasing their propulsion magnitude or propulsion direction.(Fig.~\ref{Fig_7}g).

    \item {\bf Propulsion by deformable nanoribbons in spatiotemporally oscillating fields.} Spatiotemporally structured light could induce a snake-like motion in hydrogel nanoribbons\cite{Mourran2017,Zhang2017scr,palagi2016structured} functionalized with microparticles, which would then propel (Fig.~\ref{Fig_7}h). 

    \item {\bf 3D-printed articulated microbots actuated by spatiotemporally structured light.} Recent advances in 3D printing have enabled the precise programmable control over the shape morphing and folding properties of soft stimuli-responsive composite materials.\cite{Magdanz2014,Jeon2017,Jeon2017_02,Erb2013} For example, embedded gold nanorods have been employed to enhance light absorption in temperature-responsive hydrogel composites.\cite{Mourran2017,Zhang2017scr} Printing composite hydrogels including gold nanoparticles of different sizes and aspect ratios could then enable researchers to address such hydrogels individually by exploiting different plasmonic resonances to realize artificial muscles and joints for microscale walkers and crawlers (Fig.~\ref{Fig_7}i).

\end{itemize}

\section{Conclusions}

In this perspective, we have discussed the ongoing progress towards actuating and controlling soft active matter by exploiting the different properties of light. While changing the light intensity provides a remote effortless means to adjust the speed and direction of light-activated particles, the potential of other properties of light to control soft active matter actuation has been mostly left untapped. For example, selectivity to light wavelength can enable multiple propulsion mechanisms to coexist on a single particle, e.g., by triggering exclusive light-matter interactions at different sites of the particle.\cite{Vutukuri2020,Jang2017,Wang2018,Sridhar2020}  Furthermore, light polarization and the transfer of its linear and angular momentum can enable complex combinations of translation and rotation in the propulsion of active particles without the need for any additional fuel source.\cite{tanaka2020plasmonic,andren2021microscopic,Wu2022} Finally, the use of spatiotemporally structured light can combine all of light's different degrees of freedom into one powerful tool to control the actuation of active matter systems by light in a way that is flexible, selective and adaptive, yet concomitantly easy to operate. Such level of control through light can enable active matter researchers to test theory (e.g., phase transitions, optimal navigation strategies) as well as to develop applications in energy conversion, catalysis, drug-delivery and tissue engineering taking advantage of the fact that light is a broadly available and biocompatible source of energy. Conversely, a higher degree of control of active matter with light could prove useful to develop materials and devices based on active systems that can mold the flow of light in non-conventional ways, e.g., to realize novel light sources, neuromorphic computers and, even, displays.

\newpage

\section{Acknowledgements}
M.R acknowledges Antonio Ciarlo, Giuseppe Pesce and Martin Wittmann for fruitful discussions. M.R. acknowledges funding from Marie Sklodowska-Curie Individual Fellowship (Grant No.101064381). G.V. (Giorgio Volpe) acknowledges sponsorship for this work by the US Office of Naval Research Global (Award No. N62909-18-1-2170). G.V. (Giovanni Volpe) acknowledges funding from the Horizon Europe ERC Consolidator Grant MAPEI (grant number 101001267) and the Knut and Alice Wallenberg Foundation (grant number 2019.0079).

\newpage

\clearpage

\bibliography{achemso-demo}

\end{document}